# Bibliometric analysis of the scientific production found in Scopus and Web of Science about business administration

Félix Lirio-Loli[1] [*] and William Dextre-Martínez [2]

Business School

Universidad Santiago Antúnez de Mayolo.
Ciudad Universitaria, 02001
Huaraz, Perú

aliriol@unasam.edu.pe

**ABSTRACT**

*Introduction: This study analyzes the scientific production in business administration in scientific articles based on modeling partial least squares structural equations (Partial Least Squares Structural Equation Modeling PLS-SEM) in the 2011-2020 period.*

*Methodology: The study is exploratory - descriptive and has three phases: a) Selection of keywords and search criteria; (b) Search and refinement of information; c) information analysis. A method of bibliometric review of the specific literature has been used based on the analysis of predefined indicators and completed with a qualitative content synthesis.*

*Results: A total of 167 publications were analyzed, making correlations from the year, search criteria, authors, impact factor by quartile, and by citation variables. More outstanding scientific production comes from Scopus under the search criteria ((pls AND sem) OR "partial least squares") AND (business OR management), being the figure of 4,870 scientific articles, while Web of Science accumulates 3,946 articles*

*Conclusion: There has been a progressive growth in scientific articles with the PLS-SEM technique from 2011 to 2020. Scopus, compared to WoS, presents a more significant number of scientific productions with this statistical approach. The authors who register scientific articles demonstrate a high H index; in addition, there is an important number of scientific articles with a PLS-SEM approach in universities in Malaysia that could be related to the expansion of higher education in that country, as well as in Singapore, Taiwan, and Indonesia. Finally, business administration, accounting, and economics are outstanding scientific production.*

## INTRODUCTION

Administration or management is the process of designing and maintaining environments in which individuals who collaborate in teams efficiently meet selected organizational objectives. A manager performs the functions of planning, organizing, directing, and controlling. Management applies to every organization, private or public. All around the world, business administration is essential in organizations that constitute productive and service units that generate jobs, economic dynamics, and social welfare. Management is commonly related to productivity (Drucker, 2012; Robbins & Coulter, 2017).

Business administration leads a company's management seeking its growth through the optimal use of resources and improving its processes to define success. Contingency theory points out that organizations evolve, that behaviors change over time, that the world moves very differently, that nothing remains, that everything is unpredictable. Therefore, in the study of management, it becomes crucial to divide into its functions of planning, organizing, integrating personnel, directing, and controlling, according to which knowledge can be classified (Luo et al., 2010; Witzel, 2016).

---

[1] Escuela de Administración de la Universidad Nacional Santiago Antúnez de Mayolo, Huaraz, Perú.

[*] Corresponding author: e-mail: aliriol@unaam.edu.pe (Félix Lirio-Loli).

[2] Escuela de Postgrado de la Universidad Nacional Santiago Antúnez de Mayolo, Huaraz, Perú

Because it is essential to evaluate the performance of companies, it has also become imperative to choose tools to measure the impact. Managers with financial constraints may find it more difficult to take steps to design measurement plans and collect specialized data. On the other hand, impact-oriented managers may require their organizations to adopt more vital techniques to assess the causality of the intervention. In this way, researchers could examine the diversity of entrepreneurs and investors and how their different orientations and resources explain the adoption of various tools to assess impact. Business research requires routines that use statistical and mathematical formulas to process and forecast data. The findings in business administration need transparent research methods and data analyses, justifications of the research methodology and empirical context, a clear description of the process for obtaining findings, suggestions for further research, acknowledgment of limitations, and others. Although statistical tools should use the same formulas, the practice with which they execute the data processing and visualizing, the results do not always ease the researcher. This difference in ease and practice is often related to developing new techniques that have appeared over the years (Lazzarini, 2018; Lindgreen et al., 2021; Luo et al., 2010).

**PLS-SEM, a technique applied to business administration**

The PLS-SEM is an emerging statistical procedure for structural modeling in research projects. It has the advantage of being used in the dataset with small samples without having a normal data distribution. This statistical procedure includes reliability, validity, collinearity problems, predictive relevance, and effect sizes, which need to be evaluated in advance and report the coefficient of determination and trajectory coefficients found in the research model for efficient use in scientific production. (Hair et al., 2011; Wong, 2019).

Therefore, with the increasing prominence of *Partial Least Squares - Structural Equation Modeling* (PLS-SEM) in business, researchers have increasingly turned to PLS-SEM for business research, and this statistical procedure has been adopted by science researchers as well (Hair, Sarstedt, Pieper, & Ringle, 2012; Carrion, Henseler, Ringle, & Roldan, 2016).

In business administration, the use of PLS-SEM as a statistical technique is increasingly evident (Hsu et al., 2016; Jisha & Thomas, 2016); thus, modeling in structural equations (SEM) has become an essential standard in business administration research to analyze cause-effect relationships between latent constructs (predominantly in marketing and business management research, In other words, SEM, properly applied, is a "*silver bullet*" for estimating causal models in many theoretical models and empirical data situations (Hair et al., 2011).

This research analyzes the scientific production in business administration through articles based on Partial Least Squares Structural Equation Modeling PLS-SEM from 2011 to 2020.

**METHOD**

The bibliometric method is a qualitative approach to analyze information obtained to understand and evaluate research performance (Belter, 2018; Haddow 2018). It is a non-experimental study; the variable is not introduced, controlled, or manipulated (Kumar,2018). The research was exploratory - descriptive with retrospective design. A review of the literature was carried out in other similar studies (Arana Barbier, 2020; Blanco-Encomienda & Rosillo-Díaz, 2021). Scopus and Web of Science are the tools to look for documentary sources concerning the study area.

The inclusion criteria englobe articles in the period 2011 - 2020; with no language discrimination.

The research work has three steps:

1. *Selection of keywords and* search: As the study seeks to find the scientific production with the method of *modeling structural equations* of partial least squares, the criterion was used *partial least squares structural equation modeling* next to its abbreviation *PLS-SEM. Similarly,* in the desire to focus on business administration, the *business management* criterion was used, so the search equation was defined as follows:

    ((pls AND sem) OR "partial least squares") AND (business OR management).

2. *Search and* refinement *of the* information: The query was made with the search equation in Scopus and Web of Science using the criteria by years (2011-2020), articles, number of citations per year, researchers, H index, country, affiliation, and research areas.
3. *Analysis of* the information: Tables are for the descriptive analysis with the analysis criteria.

After the quantitative analysis, the qualitative analysis leads to understanding the data obtained. This methodology fundament the primary input for developing interpretation and argumentation processes that allowed the generation of inferences and relationships thanks to the researchers' content and topic analysis (Donthu et al., 2021).

**RESULTS**

*Table 1: Number of articles per year in Scopus and WoS*

| Year | Articles in Scopus | Difference | Articles in WoS | Difference |
|---|---|---|---|---|
| 2011 | 183 | | 99 | |
| 2012 | 226 | 43 | 123 | 24 |
| 2013 | 271 | 45 | 133 | 10 |
| 2014 | 289 | 18 | 181 | 48 |
| 2015 | 294 | 5 | 215 | 34 |
| 2016 | 390 | 96 | 356 | 141 |
| 2017 | 436 | 46 | 399 | 43 |
| 2018 | 627 | 191 | 544 | 145 |
| 2019 | 886 | 259 | 813 | 269 |
| 2020 | 1,268 | 382 | 1,083 | 270 |
| Total | 4,870 | | 3,946 | |

Table 1 shows the scientific production by year in Scopus and Web of Science. It verifies that the contribution in Scopus was increasing every year, as in Web of Science. However, there are more consigned articles in Scopus.

*Table 2: Most cited articles in Scopus*

|   | Title | Total |
|---|---|---|
| 1 | PLS-SEM: Indeed, a silver bullet | **5513** |
| 2 | Using PLS path modeling in new technology research: Updated guidelines | **1134** |
| 3 | Common Beliefs and Reality About PLS: Comments on Rönkkö and Evermann (2013) | **832** |
| 4 | The Use of Partial Least Squares Structural Equation Modeling in Strategic Management Research | **640** |
| 5 | Hierarchical Latent Variable Models in PLS-SEM: Guidelines for Using Reflective-Formative Type Models | **615** |
| 6 | Using partial least squares in operations management research | **524** |
| 7 | An updated and expanded assessment of PLS-SEM in information systems research | **422** |
| 8 | Partial least squares structural equation modeling (PLS-SEM): A useful tool for family business researchers | **412** |
| 9 | An assessment of the use of (PLS-SEM) in hospitality research | **196** |
| 10 | The role of knowledge-oriented leadership in knowledge management practices and innovation | **233** |

*Table 3: Most cited articles by year in Scopus*

| Points per article | 2011 | 2012 | 2013 | 2014 | 2015 | 2016 | 2017 | 2018 | 2019 | 2020 | Total |
|---|---|---|---|---|---|---|---|---|---|---|---|
| 1 | 13 | 61 | 141 | 234 | 358 | 527 | 652 | 891 | 1110 | 1526 | **5513** |
| 2 |    |    |    |    |    | 35 | 80 | 214 | 325 | 480 | **1134** |
| 3 |    |    |    | 8 | 17 | 84 | 99 | 141 | 184 | 299 | **832** |
| 4 |    | 2 | 11 | 32 | 40 | 70 | 74 | 108 | 122 | 181 | **640** |
| 5 |    | 1 | 6 | 26 | 32 | 57 | 76 | 108 | 139 | 170 | **615** |
| 6 |    |    | 7 | 35 | 34 | 57 | 45 | 97 | 111 | 138 | **524** |
| 7 |    |    |    |    |    |    | 1 | 4 | 30 | 125 | 262 | **422** |
| 8 |    |    |    | 5 | 7 | 34 | 44 | 64 | 114 | 144 | **412** |
| 9 |    |    |    |    |    |    |    |    | 21 | 49 | 126 | **196** |
| 10 |    |    |    |    | 6 | 20 | 24 | 45 | 60 | 78 | **233** |

*Table 4: Most cited articles in Web of Science*

|   | Articles in WoS |   |
|---|---|---|
| 1 | A new criterion for assessing discriminant validity in variance-based structural equation modeling | 2088 |
| 2 | An assessment of the use of partial least squares structural equation modeling in marketing research | 1388 |
| 3 | Using PLS path modeling in new technology research: updated guidelines | 708 |
| 4 | Common Beliefs and Reality About PLS: Comments on Ronnkko and Evermann (2013) | 533 |
| 5 | Hierarchical Latent Variable Models in PLS-SEM: Guidelines for Using Reflective-Formative Type Models | 382 |
| 6 | Partial Least Squares (PLS) Structural Equation Modeling (SEM) for Building and Testing Behavioral Causal Theory | 348 |
| 7 | Using partial least squares in operations management research: A practical guideline and summary of past research | 329 |
| 8 | Consistent partial least squares path modeling | 279 |
| 9 | Testing measurement invariance of composites using partial least squares | 276 |
| 10 | Mediation analysis in partial least squares path modeling Helping researchers discuss more sophisticated models | 243 |

Table 5: Most cited articles per year in Web of Science

| Articles in WoS | 2011 | 2012 | 2013 | 2014 | 2015 | 2016 | 2017 | 2018 | 2019 | 2020 | |
|---|---|---|---|---|---|---|---|---|---|---|---|
| 1 | | | | | 6 | 88 | 146 | 305 | 589 | 954 | **2088** |
| 2 | | 15 | 37 | 61 | 113 | 174 | 170 | 218 | 282 | 318 | **1388** |
| 3 | | | | | | 30 | 41 | 133 | 214 | 290 | **708** |
| 4 | | | | 9 | 14 | 59 | 60 | 89 | 134 | 168 | **533** |
| 5 | | | 4 | 13 | 14 | 35 | 45 | 65 | 96 | 110 | **382** |
| 6 | | | | 2 | 18 | 33 | 40 | 65 | 75 | 115 | **348** |
| 7 | | | 7 | 25 | 21 | 42 | 30 | 64 | 69 | 71 | **329** |
| 8 | | | | | 1 | 29 | 21 | 52 | 76 | 100 | **279** |
| 9 | | | | | | 8 | 25 | 31 | 90 | 122 | **276** |
| 10 | | | | | | 3 | 15 | 39 | 63 | 123 | **243** |

In the case of Scopus, the most cited article is "*PLS-SEM: Indeed a silver bullet*" written by Hair et al. (2011) that accumulates a total of 5,513 citations, followed by Henseler, J (2016) con *Using PLS path modeling in new technology research: updated guidelines.*

In the case of Web of Science, the work of more citations is *A new criterion for assessing discriminant validity in variance-based structural equation modeling* of, again, Henseler et al. (2015)

With 2,088 citations, followed by *An assessment of the use of partial least squares structural equation modeling in marketing research* by Hair et al. (2012)

Table 6: Authors with the highest production in Scopus

| | Most cited author in Scopus | Articles production | H-Index |
|---|---|---|---|
| 1 | Jermsittiparsert, Kittisak | 53 | 38 |
| 2 | Ramayah T. | 35 | 44 |
| 3 | He Yong | 20 | 58 |
| 4 | Mutanga, Onisimo | 16 | 39 |
| 5 | Charoensukmongkol, Peerayuth | 15 | 15 |
| 6 | Ringle, Christian M. | 15 | 51 |
| 7 | Fongsuwan, Wanno | 14 | 4 |
| 8 | Iranmanesh, Mohammad | 14 | 25 |
| 9 | Henseler, Jörg | 13 | 34 |
| 10 | Roldán, José L. | 13 | 27 |

Table 7: Authors with the highest production in Web of Science

| | Most cited author in WoS | Articles production | H-Index |
|---|---|---|---|
| 1 | Iranmanesh, Mohammad | 44 | 23 |
| 2 | Ramayah T | 42 | 33 |
| 3 | Ringle, Christian M. | 37 | 47 |
| 4 | Sarstedt Marko | 34 | 49 |
| 5 | Ooi, Keng-Boon | 28 | 45 |
| 6 | Henseler Jorg | 25 | 32 |
| 7 | Zailani, Suhaiza | 24 | 28 |
| 8 | Roldán, José L. | 23 | 23 |
| 9 | Cheah, Jun-Hwa | 21 | 8 |
| 10 | Hair JF | 20 | 16 |

According to Table 7 and 8, the authors who stand out the most are Jermsittiparsert, Kittisak y Iranmanesh, Mohammad for Scopus, and Web of Science, respectively. Curiously Ramayah ranks second in both databases.

*Table 8: Most productive organizations in Scopus*

|    | Affiliation | Quantity |
|----|---|---|
| 1  | Universiti Sains Malaysia | 164 |
| 2  | Universiti Utara Malaysia | 161 |
| 3  | Suan Sunandha Rajabhat University | 81 |
| 4  | Universiti Malaya | 80 |
| 5  | Universidad de Sevilla | 77 |
| 6  | Universiti Teknologi MARA | 61 |
| 7  | College of Business, Universiti Utara Malaysia | 58 |
| 8  | Chinese Academy of Sciences | 57 |
| 9  | Bina Nusantara University | 53 |
| 10 | Zhejiang University | 50 |

*Table 9: Most productive organizations in Web of Science*

|    | Affiliation | Quantity |
|----|---|---|
| 1  | Universiti Sains Malaysia | 140 |
| 2  | University Of Sevilla | 127 |
| 3  | Universiti Malaya | 81 |
| 4  | Universidad De Castilla La Mancha | 68 |
| 5  | State University System Of Florida | 63 |
| 6  | Universiti Teknologi Malaysia | 60 |
| 7  | University Of Newcastle | 51 |
| 8  | Ucsi University | 49 |
| 9  | University Of Valencia | 47 |
| 10 | University System Of Georgia | 47 |

In both databases, the contribution is visible from the *Universiti Sains Malaysia* in works in favor of business administration with 164 and 140 scientific articles registered for Scopus and Web of Science, respectively.

*Table 10: Scientific production by country in Scopus*

|    | Country        | Production |
|----|----------------|------------|
| 1  | Malaysia       | 742        |
| 2  | China          | 667        |
| 3  | United States  | 516        |
| 4  | Indonesia      | 488        |
| 5  | Spain          | 451        |
| 6  | Australia      | 270        |
| 7  | United Kingdom | 244        |
| 8  | India          | 217        |
| 9  | Germany        | 183        |
| 10 | Thailand       | 174        |

*Table 11: Scientific production by country according to Web of Science*

|    | Country        | Production |
|----|----------------|------------|
| 1  | Peoples R China | 711       |
| 2  | Spain          | 587        |
| 3  | United States  | 576        |
| 4  | Malaysia       | 450        |
| 5  | Taiwan         | 337        |
| 6  | Australia      | 316        |
| 7  | England        | 271        |
| 8  | Germany        | 218        |
| 9  | South Korea    | 183        |
| 10 | Pakistan       | 160        |

The most productive country is Malaysia. The Republic of China has an actual rank in both databases. In the ranking of the ten nations, 40% are represented by Asian countries.

*Table 12: Research areas at Scopus*

|    | Research Areas                           | Record Count |
|----|------------------------------------------|--------------|
| 1  | Business, Management and Accounting      | 2354         |
| 2  | Social Sciences                          | 1141         |
| 3  | Computer Science                         | 820          |
| 4  | Engineering                              | 771          |
| 5  | Environmental Science                    | 595          |
| 6  | Agricultural and Biological Sciences     | 573          |
| 7  | Economics, econometrics and Finance      | 550          |
| 8  | Decision Sciences                        | 517          |
| 9  | Energy                                   | 267          |
| 10 | Biochemistry, Genetics and Molecular Bio | 548          |

*Table 13: Research areas in Web of Science*

|    | Research Areas                      | Record Count |
|----|-------------------------------------|--------------|
| 1  |                                     |              |
| 2  | Business Economics                  | 1944         |
| 3  | Environmental Sciences Ecology      | 550          |
| 4  | Computer Science                    | 502          |
| 5  | Engineering                         | 466          |
| 6  | Information Science Library Science | 446          |
| 7  | Social Sciences Other Topics        | 428          |
| 8  | Science Technology Other Topics     | 414          |
| 9  | Psychology                          | 195          |
| 10 | Education Educational Research      | 127          |

Table 14 shows the predomination of business administration in both databases. However, with the figure of 2,354 over 1944, Scopus stands out from Web of Science.

*Tabla 14: Relationship between articles and H index of authors in Scopus*

|                    |               | Value | Error | T value | P-value |
|--------------------|---------------|-------|-------|---------|---------|
| Ordinal by ordinal | Kendall Tau-b | ,322  | ,094  | 3,337   | ,001    |
| Valid cases        |               | 10    |       |         |         |

The symmetrical measures of agreement between the number of articles achieved and the authors' H Index in the analysis topic indicate a moderate and positive association. The correlation coefficient is 0.3224 with a p-value of 0.01 indicates a significant correlation through the works in Scopus

*Table 15: Relationship between articles and H index of authors in WoS*

|             |               | Value | Error | T Value | P-Value |
|-------------|---------------|-------|-------|---------|---------|
| Ordinal     | Kendall Tau-b | ,494  | ,250  | 1,957   | ,05037  |
| Valid Cases |               | 10    |       |         |         |

In the case of WoS, the symmetrical measures of agreement between the number of articles achieved and the authors' H index show no association because the p-value exceeds 0.05.

**DISCUSSION**

This study confirms that from 2011 to 2020, Scopus has registered an increasing number *per year* in articles linked to the PLS-SEM approach reaching the figure of 4,870. Meanwhile, WoS also registers the same increasing behavior per year, reaching 3,946 articles with the PLS-SEM model during this range of years, 81% concerning Scopus. The results clearly indicate that researchers should use Scopus and WoS to evaluate their findings. Maybe Scopus has a larger number of articles because it covers a broader range of topics than other indexes. Also, it allows free access to author and source information, including metrics, it is more accessible to the public than other databases (Abrizah et al., 2013; Bar-Ilan, 2008; Mongeon & Paul-Hus, 2016).

The most cited paper (5,513 in Scopus) is the well-known article "*PLS-SEM: Indeed a silver bullet*" written by Hair et al. (2011), and it emphasizes that PLS-SEM path modeling can be a *"silver bullet"* to estimate causal models in many situations of theoretical models and empirical data. Other researchers' criticisms of PLS-SEM are often misplaced of the true potential of the method if the objective of the analysis were better understood to choose the PLS-SEM statistical method. The article "*Using PLS path modeling in new technology research: updated guidelines*" shows that path modeling is a variance-based structural equation modeling (SEM) technique for business and social topics. It can model constructs and factors as a statistical tool for research, providing guidelines on using PLS and reporting and interpreting results (Henseler et al., 2016). In WoS, the article "*A new criterion for assessing discriminant validity in variance-based structural equation modeling*" has 2.088 cites and it was written by Henseler et al. (2015). It verifies that the evaluation of discriminant validity has become a prerequisite for analyzing the relationships between latent variables for modeling structural equations based on variances, such as partial least squares, the Fornell-Larcker criterion, and cross-load examination approaches to assessing discriminant validity in the PLS-SEM approach. "*An assessment of the use of partial least squares structural equation modeling in marketing research*" by Hair et al. (2012) has been cited 2,088 citations. It clarifies that little attention has been paid to evaluating the use of partial least squares structural equation modeling (PLS-SEM) in favor of market research, despite its growing popularity. An interpretation that it makes based the exhaustive search in the 30 best-classified marketing journals that allowed to identify 204 PLS-SEM applications published in a period of 30 years (1981 to 2010).

Regarding the researchers of more excellent production in articles of PLS-SEM stands out in Scopus *Kittisak Jermsittiparsert Dhurakij* from *Pundit University, Bangkok, Thailand* with 44 articles using PLS-SEM in Scopus; while in WoS predominates *Mohammad Iranmanesh* from *Universiti Sains Malaysia* with 44 articles with the statistical approach analyzed. Ramayah, T has second place in both databases, with 35 and 42 articles in Scopus and WoS. A high H index is detected in those who publish scientific articles with the PLS-SEM approach. He Yong is shown in Scopus with an H index of 58, and Sarstedt Marko has achieved 49 in WoS. These findings demonstrate that this index may not be a useful tool in scientific performance because other tools like the OGP are valuable tools for studying h-like index sensitivity to citations. Interestingly, the OGP finds the h-index to be the weakest performer (Mongeon & Paul-Hus, 2016).

In universities topic, *Universiti Sains Malaysia* predominates in Scopus and Web of Science public research university in Malaysia one of the oldest institutes of higher education in North Malaysia and also one of the largest in terms of enrollment in that country. Regarding the countries with the highest scientific production, those from Asia stand out with Malaysia as the leader, China, Indonesia, Thailand, Taiwan, and South Korea. In that sense, the article that reviews the conceptual and methodologies in the research on the director's instructional leadership in Malaysia to explain the construction of knowledge in a developing society (*Review of conceptual models and methodologies in research on principal instructional leadership in Malaysia: A case of knowledge construction in a developing society*) is described that in recent decades, instructional leadership has gradually gained more and more validity as a critical factor of school principals in much of the world with Malaysia being one of the countries with that case, where educational research, policy, and practice have led to educational leadership as its protagonist. However, in the evaluation made in that country, it was noted that in the last 30 years, more than 75 percent of Malaysian educational leadership studies (postgraduate theses in master's and doctorate), few had achieved their publication in journals. In addition, the authors' analysis found that most studies had used lower-order conceptual models (i.e., bivariate, direct effects) and relied heavily on descriptive and simple statistical correlation tests. This shows a cohesive Asia's countries developing economies are challenging long-held US supremacy in the global scientific community (e.g., Malasya, South Korea, India, etc.). Asian nations have become more important in science and engineering, second only to the United States in terms of articles published (Gomez et al., 2020; Xie et al., 2014).

The lack of consistent results within the studies database was attributed to limitations in research design and quality. Even so, limitations in the research models and methods employed in these studies suggest the need for more vital methodological training before Malaysian academics can achieve the goal of contributing helpful knowledge to the local and global knowledge base (Hallinger et al., 2018).

Business administration is the research area that brings the highest concentration of scientific articles with the common denominator of PLS-SEM. A finding by Dash et al. (2021) that supports the greater use of PLS-SEM over CB-SEM being the *average variance extracted* (AVE) and composite reliability values (CR) the most used techniques in this method, which indicates better reliability and construct validity in research. It is also pertinent to carry out bibliometrics with all the currents of business administration to know the scientific production of each of more topics from business administration (human resources, marketing, strategic planning) to compare them with the findings of the present study.

## CONCLUSION

The study shows annual growth in scientific articles under the PLS-SEM technique in Scopus in Web of Science for 2011 - 2020. Scopus has a tremendous amount of scientific production with this statistical approach.

The most cited articles belong to PLS-SEM methodology as a robust and advanced technique for prediction in econometric models of multiple equations. In this way, the quantitative analysis PLS-SEM has gained a presence in most top academic journals. It shows that the academic community is developing work to apply PLS-SEM correctly, even though the technique is still considered an emerging multivariate data analysis method, and researchers are still exploring PLS-SEM best practices.

The authors who register scientific articles considering the PLS-SEM technique show a high H index, with a significant correlation between scientific production and the H index in Scopus (p-value = 0.01), but not in WoS (p-value = 0.05037).

A vital production of scientific articles with the PLS-SEM was written in Universities in Malaysia, an example of *higher education expansion* country, as well as in Singapore, Taiwan, and Indonesia that show advances in the production of research with the PLS-SEM approach.

Business administration, accounting, and economics produce more scientific production through the PLS-SEM technique.